\documentclass[12pt,reqno]{article}
\usepackage{amsfonts}
\usepackage{amsmath}
\usepackage{amsbsy}

\usepackage{amssymb,latexsym}
\numberwithin{equation}{section}

\begin{document}
 \allowdisplaybreaks[1]
\title{Decoupling Structure of the Principal Sigma Model-Maxwell Interactions}
\author{Nejat T. Y$\i$lmaz\\
Department of Mathematics
and Computer Science,\\
\c{C}ankaya University,\\
\"{O}\u{g}retmenler Cad. No:14,\quad  06530,\\
 Balgat, Ankara, Turkey.\\
          \texttt{ntyilmaz@cankaya.edu.tr}}
\maketitle
\begin{abstract}
The principal sigma model and Abelian gauge fields coupling is
studied. By expressing the first-order formulation of the gauge
field equations an implicit on-shell scalar-gauge field decoupling
structure is revealed. It is also shown that due to this
decoupling structure the scalars of the theory belong to the pure
sigma model and the gauge fields sector consists of a number of
coupled Maxwell theories with currents partially induced by the
scalars.
\end{abstract}

\section{Introduction}
The scalar sectors of the supersymmetric field theories especially
the supergravity theories which are the theories that govern the
massless sector low energy background coupling of the relative
superstring theories \cite{kiritsis} can be formulated as
principal sigma models or non-linear sigma models whose target
spaces are group manifolds or coset spaces. In particular a great
majority of the supergravity scalar sectors are constructed as
symmetric space sigma models
\cite{sm1,sm2,sm3,nej1,nej2,nej3,sssm1}. When the Abelian
(Maxwell) vector multiplets are coupled to the graviton multiplets
in these theories the scalar sector which has the non-linear sigma
model interaction with in itself is coupled to the Abelian gauge
fields through a kinetic term in the Lagrangian
\cite{julia1,julia2,ker1,ker2}.

In this work, we will focus on the gauge field equations of the
principal sigma model and the Abelian gauge field couplings
mentioned above. Bearing in mind that the non-linear sigma model
can be obtained from the principal sigma model by imposing extra
restrictions for the sake of generality we will consider the
generic form of the principal sigma model as the non-linear
interaction of the scalars. We will simply show that the gauge
field equations can be locally integrated so that they can be
expressed as first-order equations containing arbitrary locally
exact differential forms. The first-order form of the gauge field
equations will be used to show that there exists a one-sided
decoupling between the scalars and the gauge fields in a sense
that the scalar field equations do not contain the gauge fields in
them whereas the scalars enter as sources in the gauge field
equations. Thus the scalar solution space of the coupled theory
coincides with the general solution space of the pure sigma model.
Furthermore we will also discuss that this hidden on-shell
scalar-matter decoupling results in a number of coupled Maxwell
theories with sources whose currents contain the general solutions
of the principal sigma model which is completely decoupled from
the Maxwell sector. Therefore we will show that when the general
solutions of the pure principal sigma model are obtained and when
one fixes the sector of the solution space of the coupled theory
by fixing the field dependence or the independence of the locally
exact differential forms appearing in the first-order gauge field
equations one may determine the currents of the coupled Maxwell
fields. In this respect one may solve the gauge fields from the
Maxwell sector field equations. Consequently the solution space of
the scalar-matter coupling can be entirely generated by the
general solutions of the pure principal sigma model and the
arbitrary choice of the locally exact differential forms. This
fact is a consequence of the solution methodology which is based
on the implicit on-shell decoupling between the matter fields and
the scalar sector which provides current sources to the former.
\section{Hidden Decoupling Structure of the Gauge Fields and Their Sources}
In a $D$-dimensional spacetime $M$ the Lagrangian which gives the
inhomogeneous Maxwell equations can be given as
\begin{equation}\label{de1}
 {\mathcal{L}}=-\frac{1}{2}dA\wedge
 \ast dA-A\wedge\ast J,
\end{equation}
where $A$ is the $U(1)$ electromagnetic gauge potential one-form
and $F=dA$ is the field strength of it. Also in the units where
the speed of light is unity the current one-form in a local
coordinate basis $\{dt,dx^{a}\}$ is
\begin{equation}\label{de2}
 J=-\rho dt+\mathbf{J}_{a}dx^{a},
\end{equation}
where in the temporal component $\rho$ is the charge density and
the spatial components $\mathbf{J}_{a}$ are the current densities.
The Lagrangian \eqref{de1} defines a media in which the charge
density and the currents are not influenced by the electromagnetic
field. The current one-form is predetermined and static that is to
say although it acts as a source for the electromagnetic field it
does not interact with it dynamically. From \eqref{de1} the
inhomogeneous Maxwell equations read
\begin{equation}\label{de2.5}
 d\ast F=-\ast J.
\end{equation}
In this section we will consider the coupling of $N$ $U(1)$ gauge
field one-forms $A^{i}$ to the principal sigma model whose target
space is a group manifold $G$. The sigma model Lagrangian can be
given as
\begin{equation}\label{de3}
 {\mathcal{L}}=\frac{1}{2}\, tr(\ast dg^{-1}\wedge
 dg).
\end{equation}
Here we take a differentiable map
\begin{equation}\label{de4}
 h: M\longrightarrow G,
\end{equation}
we also consider a representation $f$ of $G$ in $Gl(N,\Bbb{R})$
\begin{equation}\label{de5}
 f: G\longrightarrow Gl(N,\Bbb{R}),
\end{equation}
which may be taken as a differentiable homomorphism. Then the map
$g$ can be given as
\begin{equation}\label{de6}
 g=f\circ h: M\longrightarrow Gl(N,\Bbb{R}),
\end{equation}
which is a matrix-valued function on $M$
\begin{equation}\label{de7}
g(p)=\left(\begin{array}{ccc}
  \varphi^{11}(p) & \varphi^{12}(p) & \cdots \\
  \varphi^{21}(p) & \varphi^{22}(p) & \cdots \\
  \vdots & \vdots & \vdots \\
\end{array}\right),
\end{equation}
$\forall p\in M$. Due to the presence of the scalar fields
$\{\varphi^{ij}\}$ the theory can be considered to be a scalar
field theory. In \eqref{de3} we have a matrix multiplication with
the wedge product used between the components and the trace is
over the representation chosen in \eqref{de5}. Depending on the
restrictions on the sigma model and the nature of $G$ the scalars
in \eqref{de7} can be all independent or not. This model also
covers the non-linear (coset) sigma models \cite{westsugra,tanii}
and in particular the symmetric space sigma models
\cite{sm1,sm2,sm3} which shape the scalar sectors of the
supergravities thus the low energy effective string theories. The
construction of the Lagrangian of the symmetric space sigma model
in which an internal metric substitutes the map $g$ can be found
in \cite{nej1,nej2,nej3}.

By generalizing the supersymmetric coupling of the supergravity
matter multiplets with the graviton multiplets
\cite{julia1,julia2,ker1,ker2,sssugradivdim} we can write down the
coupling of $N$ $U(1)$ gauge field one-forms $A^{i}$ to the
principal sigma model whose target space is a group manifold as
\begin{equation}\label{de8}
 \mathcal{L}_{tot}=\frac{1}{2}tr( \ast dg^{-1}\wedge dg)
 -\frac{1}{2}\ast F^{T}g\wedge
 F,
\end{equation}
where we define the column vector $F$ whose components are $F^{i}$
thus the coupling term can be explicitly written as
\begin{equation}\label{de9}
-\frac{1}{2}\ast F^{T}g\wedge
 F =-\frac{1}{2}g^{i}_{\:\:\:j} \ast F_{i}\wedge
 F^{j}.
\end{equation}
 Now if we
vary the Lagrangian \eqref{de8} with respect to $A^{i}$ we find
the corresponding field equations as
\begin{equation}\label{de10}
d(\mathcal{T}^{i}_{\:\:\:k}\ast F_{i})=0,
\end{equation}
where
\begin{equation}\label{de11}
\mathcal{T}^{i}_{\:\:\:k}=g^{i}_{\:\:\:k}+(g^{T})^{i}_{\:\:\: k}.
\end{equation}
The expression \eqref{de10} defines a closed form. Since locally
any closed form is equal to an exact form we can integrate
\eqref{de10} to write
\begin{equation}\label{de12}
\mathcal{T}^{i}_{\:\:\:k}\ast F_{i}=dC_{k},
\end{equation}
where $\{C_{k}\}$ are arbitrary $N$ $(D-3)$-forms on $M$. They may
be chosen to depend on the fields $\{\varphi^{ij},A^{i}\}$ or they
may be fixed. The former case is the most general one. The general
solutions of the field equations of the Lagrangian \eqref{de8}
must satisfy \eqref{de12} on-shell in which one can freely change
the set of $(D-3)$-forms $\{C_{k}\}$ which may be functions of
$\{\varphi^{ij},A^{i}\}$ or not. To generate the entire solution
space one can first choose a set of field-dependent or
field-independent $\{C_{k}\}$ and solve the system of field
equations of \eqref{de8} to find the corresponding solutions then
one can repeat this procedure for a different set of $\{C_{k}\}$.
In this respect the field-independent or the fixed $(D-3)$-forms
$\{C_{k}\}$ can be considered as the integration constants coming
from the reduction of the degree of the gauge field equations
which are second-order differential equations. Now we will follow
the methodology described above to show that there exists an
on-shell decoupling between the scalars and the gauge fields. Let
us first take a look at the more restricted case of
field-independent $\{C_{k}\}$ and let us fix a set of
field-independent $\{C_{k}\}$. In this case if we take the partial
derivative of both sides of \eqref{de12} with respect to the
scalar fields $\{\varphi^{ml}\}$ we immediately see that
\begin{equation}\label{de13}
\frac{\partial\mathcal{T}^{i}_{\:\:\:k}}{\partial\varphi^{ml}}\ast
F_{i}=0,
\end{equation}
since the right hand side of \eqref{de12} is a fixed $(D-2)$-form
and it does not depend on the scalar fields $\{\varphi^{ml}\}$.
These conditions must be satisfied on-shell by a sector of the
solution space of $\{\varphi^{ml},A^{i}\}$ which is restricted to
the condition of choosing fixed $\{C_{k}\}$. After fixing the set
of $(D-3)$-forms $\{C_{k}\}$ if we use \eqref{de12} in \eqref{de8}
we can obtain the on-shell Lagrangian as
\begin{equation}\label{de14}
 \mathcal{L}_{tot}=\frac{1}{2}tr( \ast dg^{-1}\wedge dg)
 -\frac{1}{2}dC_{j}\wedge
 F^{j}.
\end{equation}
Since we have made use of the gauge field equations varying this
on-shell Lagrangian with respect to $A^{i}$ yields an identity. On
the other hand by varying the above Lagrangian with respect to the
scalars to find the solutions which satisfy \eqref{de12} for a
chosen fixed set of $\{C_{k}\}$ we find that the scalar field
equations are the same ones which can be obtained directly from
the pure sigma model Lagrangian \eqref{de3} since the coupling
part in \eqref{de14} does not depend on the scalars. This result
can also be obtained by directly varying \eqref{de8} and by using
the conditions \eqref{de13} which are satisfied by the sector of
the general solutions of the theory which we have restricted
ourselves in by fixing $\{C_{k}\}$. The same result would be
obtained if one chooses another field-independent set of
$\{C_{k}\}$. Thus we observe that the scalar solutions of a
sub-sector of the scalar-gauge field theory defined by the
coupling Lagrangian \eqref{de8} are the same with the general
solution space of the pure principal sigma model. This is a
consequence of the local integration given in \eqref{de12} and
fixing the arbitrary $\{C_{k}\}$.

On the other hand a similar result with a minor difference can
also be derived for the rest of the solution space of the theory.
If we consider the most general case of field-dependent
differential forms $\{C_{k}(\varphi^{ml},A^{i})\}$ and use
\eqref{de12} in the Lagrangian \eqref{de8} we get
\begin{equation}\label{de14.5}
 \mathcal{L}_{tot}=\frac{1}{2}tr( \ast dg^{-1}\wedge dg)
 -\frac{1}{2}dC_{j}(\varphi^{ml},A^{i})\wedge
 F^{j}.
\end{equation}
This on-shell Lagrangian gives us the same decoupling conditions
discussed above for the restricted sub-sector case. To see this we
should realize that the second term in the Lagrangian
\eqref{de14.5} which is written in an on-shell form is a closed
differential form. Locally any closed differential form is an
exact one. Thus the second term in the Lagrangian \eqref{de14.5}
can be written as an exact differential form as
\begin{equation}\label{de14.6}
 -\frac{1}{2}dC_{j}(\varphi^{ml},A^{i})\wedge
 F^{j}=dB(\varphi^{ml},A^{i}).
\end{equation}
In fact we may simply calculate $B(\varphi^{ml},A^{i})$ as
\begin{equation}\label{de14.65}
 B(\varphi^{ml},A^{i})=-\frac{1}{2}C_{j}(\varphi^{ml},A^{i})\wedge
 F^{j}.
\end{equation}
 Thus the Lagrangian \eqref{de14.5} becomes
\begin{equation}\label{de14.7}
 \mathcal{L}_{tot}=\frac{1}{2}tr( \ast dg^{-1}\wedge dg)
 +d(-\frac{1}{2}C_{j}(\varphi^{ml},A^{i})\wedge
 F^{j}).
\end{equation}
If one varies the above Lagrangian one immediately sees that the
second term does not contribute to the field equations as by using
the Stoke's theorem we have
\begin{equation}\label{de14.75}
\int\limits_{M}d\delta B(\varphi^{ml},A^{i})=\int\limits_{\partial
M}\delta B(\varphi^{ml},A^{i}).
\end{equation}
If $M$ does not posses a boundary the right hand side of the above
equation is automatically zero whereas if it has a boundary then
the usual variation principles demand that the variation of the
fields on the boundary are chosen to be zero in which case again
the right hand side of \eqref{de14.75} becomes zero. Therefore we
conclude that the on-shell Lagrangian \eqref{de14.5} which is
responsible for the most general structure of the solution space
gives us the scalar field equations which are the same with the
pure sigma model field equations since the coupling part in
\eqref{de14.5} does not contribute to the scalar field equations
at all as we have proven as an on-shell condition above. Also like
we have already encountered for the non-field-dependent
$\{C_{k}\}$ sub-sector case varying \eqref{de14.5} does not give
any information (which may also mean an identity) about the gauge
fields $A^{i}$ since we have already made use of their field
equations in writing it.

In summary, we have shown that the solutions of the theory must
obey \eqref{de12} for arbitrary right hand sides. If one
determines the right hand sides in \eqref{de12} one restricts him
or herself to a layer of solutions. In this case \eqref{de12}
which are the descendants of the gauge field equations become
on-shell conditions. We have proven that if we use these
first-order field equations which are on-shell conditions back in
the general Lagrangian \eqref{de8} then the scalars of this
particular layer of solutions do not have gauge fields in their
field equations. Therefore the scalars of this particular layer
belong to the general solutions of the pure principal sigma model.
By running the right hand sides in \eqref{de12} over the entire
local field-dependent exact $(D-2)$-forms one can generalize this
result to the whole solution space. Thus in this way we have
proven that the entire scalar solutions of the coupled theory
coincide with the pure sigma model solution space. Now from
\eqref{de12} we can write down the gauge field strengths as
\begin{equation}\label{de15}
F_{l}=(-1)^{s}(\mathcal{T}^{-1})^{k}_{\:\:\:l}\ast
dC_{k}(\varphi^{mn},A^{i}),
\end{equation}
where $s$ is the signature of the spacetime. We observe that if we
consider the sector of the solution space generated by the
field-independent $\{C_{k}\}$ then after obtaining the general
solutions of the pure principal sigma model and after choosing a
fixed set $\{C_{k}\}$ one can use these in \eqref{de15} to find
the corresponding $U(1)$ gauge field strengths. Also for the more
general field-dependent case of $\{C_{k}(\varphi^{ml},A^{i})\}$
one again obtains the general solutions of the pure principal
sigma model then one inserts these solutions in \eqref{de15} to
solve for the gauge fields. We can say that in general we have a
partial decoupling between the scalars and the gauge fields. The
scalars are not affected by the presence of the gauge fields on
the other hand as we will show next they act as sources for the
gauge fields. Now after multiplying by the Hodge star operator if
we take the exterior derivative of both sides of \eqref{de15} we
obtain
\begin{equation}\label{de16}
d\ast F_{l}=(d\mathcal{T}^{-1})^{k}_{\:\:\:l}\wedge
dC_{k}(\varphi^{mn},A^{i}).
\end{equation}
When we compare this result with the inhomogeneous Maxwell
equations \eqref{de2.5} we observe that the current one-forms
become
\begin{equation}\label{de17}
J_{l}=(-1)^{(D+s)}\ast((d\mathcal{T}^{-1})^{k}_{\:\:\:l}\wedge
dC_{k}(\varphi^{mn},A^{i})).
\end{equation}
One can furthermore verify that the currents in \eqref{de17} obey
the current conservation law
\begin{equation}\label{de18}
d\ast J_{l}=0,
\end{equation}
which guarantees that the equations \eqref{de16} have solutions.
In a local moving co-frame field $\{e^{\alpha}\}$ on the spacetime
$M$ if we introduce the components of the one-forms
$(d\mathcal{T}^{-1})^{k}_{\:\:\:l}$, the $(D-2)$-forms
$dC_{k}(\varphi^{ml},A^{i})$, and the one-forms $J_{l}$ as
\begin{subequations}\label{de19}
\begin{align}
(d\mathcal{T}^{-1})^{k}_{\:\:\:l}&=(\mathcal{T}^{-1k}_{l})_{\alpha}e^{\alpha},\notag\\
\notag\\
dC_{k}(\varphi^{ml},A^{i})&=\frac{1}{(D-2)!}(\mathcal{C}_{k})_{\alpha_{1}\cdots\alpha_{(D-2)}}e^{\alpha_{1}\cdots
\alpha_{(D-2)}},\notag\\
 \notag\\
J_{l}&=\mathcal{J}_{l\beta}e^{\beta},\tag{\ref{de19}}
\end{align}
\end{subequations}
from \eqref{de17} we can calculate the components of the current
one-forms as
\begin{equation}\label{de20}
\mathcal{J}_{l\beta}=\frac{(-1)^{s}\sqrt{|detH|}}{(D-2)!}(\mathcal{T}^{-1k}_{l})_{\alpha}
(\mathcal{C}_{k})_{\alpha_{1}\cdots\alpha_{(D-2)}}H^{\alpha_{1}\beta_{1}}\cdots
H^{\alpha\beta_{(D-1)}}\varepsilon_{\beta_{1}\cdots\beta_{(D-1)}\beta},
\end{equation}
where we have introduced the metric $H$ on $M$ and the Levi-Civita
symbol $\varepsilon$. In \eqref{de20} $
\alpha,\beta,\alpha_{i},\beta_{j}=1,\cdots,D$. Also for
$\alpha_{i}$ $i=1,\cdots,D-2$ and for $\beta_{j}$
$j=1,\cdots,D-1$.

If one cancels an exterior derivative (performs integration) on
both sides of \eqref{de16} one finds
\begin{equation}\label{de21}
\ast
F_{l}=(\mathcal{T}^{-1})^{k}_{\:\:\:l}dC_{k}(\varphi^{mn},A^{i})+dC^{\prime}_{l}(\varphi^{mn},A^{i}).
\end{equation}
Now if we compare this with the first-order equations \eqref{de12}
originating from \eqref{de8} we see that our model is a sub-sector
of the one defined in \eqref{de21} with the choice of
$dC^{\prime}_{l}(\varphi^{mn},A^{i})=0$. One may also inspect
which scalar-gauge field coupling kinetic term would result in a
first-order formulation of gauge fields in the form \eqref{de21}.

We should finally state that by using the local first-order
formulation \eqref{de12} of the gauge field equations of
\eqref{de8} we have shown that there exists a decoupling between
the $U(1)$ gauge fields and the scalar fields of the theory. The
scalars are proven to be the general solutions of the pure
principal sigma model and they generate current sources for the
gauge fields as can be explicitly seen in \eqref{de16}. For the
sub-sector of the solution space which is generated by the
field-independent $\{C_{k}\}$ we end up with a decoupled set of
$N$ non-interacting Maxwell theories with prescribed and known
currents whose sources are predetermined by the principal sigma
model scalar fields. These currents interact with each other via
the sigma model which is completely decoupled from the Maxwell
sector and they define a media which does not interact with the
Maxwell fields. For this sub-sector the $N$ decoupled and
non-interacting Maxwell theories can alternatively be formulated
by the Lagrangian
\begin{equation}\label{de22}
 {\mathcal{L}}^{\prime}=\sum\limits_{i=1}^{N}\bigg(-\frac{1}{2}dA^{i}\wedge
 \ast dA^{i}+A^{i}\wedge (d\mathcal{T}^{-1})^{k}_{\:\:\:i}\wedge
 dC_{k}\bigg).
\end{equation}
We see that for a single scalar field the above Lagrangian drops
to be the ordinary Maxwell Lagrangian with a known current
one-form. We also realize that the Maxwell fields in this
restricted case are coupled to each other only by means of
integration constants. Each of these decoupled Maxwell theories is
an embedding into the ordinary Maxwell theory with known sources.
We should remark that \eqref{de22} must only be used to
derive\footnote{By integrating the field equations to first-order
and by choosing $dC^{\prime}_{l}=0$.} the field equations of the
gauge fields which belong to a restricted sector of the solution
space generated by fixing $\{C_{k}\}$ and in this case the field
equations of the scalars must again be derived from \eqref{de3}.
On the other hand for the most general solution space elements
generated by choosing field-dependent
$\{C_{k}(\varphi^{ml},A^{i})\}$ we can again adopt the general
solutions of the pure sigma model as the general scalar solutions
of our theory however in this case we have $N$ coupled Maxwell
theories whose potentials may enter in the currents. For the
sub-sector generated by $\{C_{k}(\varphi^{ml})\}$ which are not
dependent on the gauge fields we have a similar situation of
decoupled Maxwell theories with prescribed currents discussed
above.

Finally before concluding we should summarize the local general
solution methodology of the principal sigma model and $U(1)$ gauge
fields coupling defined in \eqref{de8}. The method has three steps
first find the general scalar field solutions of the pure
principal sigma model, secondly define field-dependent or
field-independent $\{C_{k}\}$ and use the general scalar solutions
in them then insert these in \eqref{de15} to solve the
corresponding gauge field strengths. By combining the general pure
principal sigma model solutions with a different set of
$\{C_{k}\}$ each time one can generate the entire set of
solutions. Such a solution methodology of the general solution
space of \eqref{de8} is a consequence of the first-order
formulation of the gauge field equations in \eqref{de12} and the
on-shell decoupling between the scalars and the gauge fields which
we have described in detail for both the restricted sector of the
solution space and for the entire solution space in its most
general generation process. From the scalars point of view there
is a complete decoupling so that the scalars of the coupled theory
coincide with the pure sigma model solution space. However the
gauge fields are not decoupled from the scalars since we have
shown that the scalars act in the current part of the gauge field
equations.
\section{Conclusion}
By integrating the gauge field equations of the principal sigma
model and Abelian gauge field coupling Lagrangian we have
expressed these equations in a first-order form. Then we have
shown that when these first-order and on-shell expressions which
contain local $(D-3)$-forms in them are used in the scalar-matter
Lagrangian one reveals an implicit decoupling between the scalars
and the matter gauge fields. We have proven that this decoupling
structure exists for the entire solution space. Therefore it is
shown that the scalar solutions of the coupled theory are the
general solutions of the pure principal sigma model. We should
state here that a similar result is derived in \cite{consist2} for
the heterotic string. However in that work the decoupling occurs
due to the existence of a dilatonic field and one derives the
decoupling structure of the coset scalars from the field equation
of the dilaton. On the other hand in the present work we prove
that such a decoupling exists for a generic principal sigma model
with an arbitrary number of Abelian gauge field couplings.

We have also mentioned that the on-shell hidden scalar-matter
decoupling mentioned above generates a theory composed of the pure
sigma model and a number of coupled Maxwell theories with sources
induced by the scalars. In particular we have discriminated a
sub-sector of the general theory generated by the
field-independent integration constants. This sub-sector contains
a number of separate and dynamically non-interacting Maxwell
theories whose current sources are drawn from the general scalar
solutions of the pure principal sigma model and the integration
constants of the first-order formulation.

In this work, we prove that the scalars of the coupled theory are
not affected by the presence of the $U(1)$ gauge fields however
they take role in the currents of the gauge fields. Thus one may
call such a coupling between the scalars and the gauge fields a
one-sided or a one and a half coupling. We also see that for the
restricted sub-sector of the theory which is obtained by fixing
the integration constants the currents of the electromagnetic
fields interact with in each other via the sigma model but they do
not interact with the corresponding gauge fields. Thus for this
sub-sector the currents form up a subset of the general currents
of the prescribed current-Maxwell theory. In this sub-sector the
scalar-gauge field decoupling induces another decoupling scheme
also among the gauge fields and one obtains $N$ non-interacting
Maxwell theories whose sources come from the pure principal sigma
model. For each $U(1)$ gauge field we have an embedded sector of
the general predetermined current-Maxwell theory which has the
most general form of currents. In our case these currents are
restricted to the solutions of the pure principal sigma model. The
$N$ $U(1)$ gauge fields in this case probe each other only by the
presence of the integration constants however this does not
correspond to a dynamic coupling. The coupling among the gauge
fields occur only at the level of determining the integration
constants from the boundary conditions. Since we have a static and
unchanged current structure which emerges from the pure principal
sigma model where the currents interact with each other and since
these currents are not affected by the presence of the gauge
fields this sub-sector of the general theory defines a theory of a
static media with $N$ non-dynamically interacting Maxwell fields.
One can inspect the other sectors of the general theory based on
various choice of field dependent $\{C_{k}\}$ which define dynamic
media with current-gauge field interactions. For example one can
study the sectors that result in gauge field equations that define
dynamic media with specially defined conducting properties
\cite{thring}. The analysis of this work can furthermore be
extended on another scalar-gauge field coupling which might have
broader dynamic media sub-sectors namely on the gauged sigma
model.

In conclusion, we may state that the decoupling structure studied
in this work is an important remark on the solutions of the
scalar-gauge field interactions which form up a basic sector of
the Bosonic dynamics of the supergravities and the effective
string theories. Therefore the scalar-gauge field decoupling
revealed here points out a simplification in seeking solutions to
these theories.

\end{document}